\documentclass[twecolumn]{pasj00}
\draft
\SetRunningHead{}{}
\Received{2002 September 2}
\Accepted{2003 January 31}
\Published{}

\begin{document}

\title{X-Ray and Gamma-Ray Emission from \\
       the PSR 1259-63/Be Star System}
\author{Kenji \textsc{Murata}, Hidenori \textsc{Tamaki},
        Hideki \textsc{Maki}, and Noriaki \textsc{Shibazaki}}
\affil{Department of Physics, Rikkyo University,
       Nishi-Ikebukuro, Tokyo 171-8501, Japan}
\email{maki@hel.rikkyo.ne.jp}


\maketitle

\begin{abstract}
PSR 1259-63 is a radio pulsar orbiting a Be star
in a highly eccentric orbit.
Soft and hard X-rays are observed from this binary system.
We apply the shock powered emission model to this system.
The collision of the pulsar and Be star winds forms a shock,
which accelerates electrons and positrons to the relativistic energies.
We derive the energy distribution of relativistic electrons and positrons
as a function of the distance from the shock in the pulsar nebula.
We calculate the X-rays and  $\gamma$-rays emitted from
the relativistic electrons and positrons
in the nebula at various orbital phases,
taking into account the Klein-Nishina effect fully.
The shock powered emission model can explain
the observed X-ray properties approximately.
We obtain from the comparison with observations
that a fraction of $\sim 0.1$ of the pulsar spin-down luminosity
should be transformed into the relativistic electrons and positrons.
We find that the magnetization parameter of the pulsar wind,
the ratio of the Poynting flux to the kinetic energy flux,
is $\sim 0.1$ immediately upstream
of the termination shock of the pulsar wind,
and may decrease with distance from the pulsar.
We predict the flux of $10~\mathrm{MeV}\!-\!100~\mathrm{GeV}$
$\gamma$-rays which may be nearly equal to the detection
threshold in the future projects.
\end{abstract}

\section{Introduction}

PSR 1259-63 is a 48 ms radio pulsar orbiting a Be star
in a 3.4 yr orbit with eccentricity of 0.86
(\cite{johns92a}; \cite{manche95}).
The binary companion is a 10th magnitude B2e star SS 2883,
whose mass and radius are estimated to be $\sim 10\ \MO$
and $\sim~6\!-\!10\ \RO$,
respectively \citep{johns92b}.
The distance may be $\sim 2\ \mathrm{kpc}$
\citep{tay93, johns94}.

The PSR 1259-63/SS 2883 system has been observed
in the X-ray and $\gamma$-ray bands ranging
from $1\ \mathrm{keV}$ to $\sim 1\ \mathrm{TeV}$
\citep{comin94, grei95, kas95, grov95, tav96, hira96, hira99, kawa02}.
The X-ray spectra are well represented by a power law function.
The X-ray luminosity in the $1\!-\!10\ \mathrm{keV}$ band is
$\sim 10^{33}\!-\!10^{34}\ \mathrm{erg \cdot s^{-1}}$.
The spectral slope and luminosity of X-rays vary with orbital phase.
The pulsation is absent in the X-ray intensity.
Only upper limits are reported for the high energy
$\gamma$-ray emission above 1 MeV including TeV $\gamma$-rays
(\cite{tav96}; \cite{kawa02}).

\citet{tav94} and \citet{tav97}
developed the shock powered emission model after they examined
a few possibilities for the high energy emission from
the PSR 1259-63/SS 2883 system.
The pulsar wind collides with the stellar wind ejected
from the Be star, forming a shock.
The shock accelerates the pulsar wind particles,
electrons and positrons, to the relativistic high energies
(\cite{hoshi92}; \cite{hoshi02}).
The relativistic electrons and positrons in the nebula radiate
$\mathrm{keV\!-\!MeV}$ photons through the synchrotron emission
and $\mathrm{GeV\!-\!TeV}$ photons through the inverse Compton
scattering of the photons from the Be star.
Model calculations for the X-ray spectral and luminosity variation
with orbital phase reproduced observations.
Note, however, that the approximation they adopted for
the Klein-Nishina effect is not appropriate especially
to see the spectral properties.
Some conclusions in \citet{tav97} may be modified
if we include the Klein-Nishina effect correctly in calculations
\citep{kirk1999a}.

\citet{kirk1999a} calculated light curves
for the hard $\gamma$-ray emission at energies up to several TeV,
including the Klein-Nishina effect and adopting several
approximations for the inverse Compton scattering process.
They use the continuous approximation for the electron energy loss
and the delta function approximation for
the distribution of target and scattered photons.
They find the light curve of hard $\gamma$-rays especially
at 100~GeV becomes asymmetric with respect to periastron
since the inverse Compton emissivity depends on
the scattering angle (the angle between the line of sight
and the vector connecting the stars).
They also show that we can expect the significantly larger
flux of hard $\gamma$-rays when the inverse Compton cooling dominates
over the synchrotron cooling.
We should note that the upper limit on the $\sim\mathrm{TeV}$
$\gamma$-ray flux, recently set by the CANGAROO collaboration,
is almost equal to the flux they predict.

We study the shock powered emission model for
the PSR 1259-63/Be Star System.
We calculate the X-rays and $\gamma$-rays emitted from
the shock accelerated electrons and positrons at various orbital phases,
taking into account the Klein-Nishina effect and
the distributions for the target and scattered photons.

This paper is organized as follows.
We describe a pulsar wind and a termination shock briefly
in section \ref{sec:wind_shock}.
We estimate cooling times of relativistic electrons and positrons
injected into the downstream region of the shock
in section \ref{sec:cooling_times}.
We follow the evolution of particle distribution
in the shock downstream region and calculate the X-rays and
$\gamma$-rays emitted from the nebula at
various orbital phases in section \ref{sec:nebula_emission}.
Comparing with X-ray observations,
we set constraints on the property of a pulsar wind
in section \ref{sec:comparison}.
In the last section we give a few concluding remarks.

\section{A Pulsar Wind and a Shock}
\label{sec:wind_shock}

Pulsars lose a large fraction of their rotational energy
in the form of relativistic MHD winds.
The pulsar wind carries $\mathrm{e^+e^-}$ pairs,
possibly together with
a small admixture of ions \citep{hoshi92}
and also conveys electromagnetic energy.
The energy flux of the pulsar wind at the distance $r$
from the pulsar can be expressed as
\begin{equation}
	F_\mathrm{w} =
		\frac{\dot{E}_{\mathrm{rot}}}{\Omega_\mathrm{w} r^2},
\label{eq:Fw1}
\end{equation}
where $\dot{E}_\mathrm{rot}$ is the spin-down luminosity of the pulsar 
and $\Omega_\mathrm{w}$ the solid angle of the wind.
We introduce the magnetization parameter $\sigma$,
defined by the ratio of the Poynting flux to the kinetic energy flux,
\begin{equation}
	\sigma = \frac{{B_1}^2}{4\pi\gamma_\mathrm{w}\rho c^2},
\label{eq:sigm}
\end{equation}
where $\gamma_\mathrm{w}$ and $\rho$ are respectively
the Lorentz factor and mass density of the pulsar wind,
$B_1$ is the magnetic field, and $c$ is the speed of light.
Then, the energy flux can also be written as
\begin{equation}
	F_\mathrm{w} = \gamma_\mathrm{w}\rho c^2(1+\sigma)c.
\label{eq:Fw2}
\end{equation}
Combining equations (\ref{eq:Fw1}), (\ref{eq:sigm}) and (\ref{eq:Fw2}),
we have
\begin{equation}
	B_1 = \left(\frac{\sigma}{1+\sigma}\right)^{1/2}
			\left(\frac{4\pi}{\Omega_\mathrm{w}}\right)^{1/2}
			\left(\frac{\dot{E}_\mathrm{rot}}{c}\right)^{1/2}
			\frac{1}{r}.
\label{eq:mag1}
\end{equation}
The strength of magnetic field decreases in inverse
proportion to the distance from the pulsar.

The pulsar wind collides with the stellar wind ejected from
the Be star, forming a shock.
We estimate the shock location from the balance
in dynamical pressure between the pulsar and stellar winds,
\begin{equation}
	\frac{\dot{E}_\mathrm{rot}}{\Omega_\mathrm{w}r^2 c}
	= \frac{\dot{M}_\mathrm{B}v_\mathrm{B}}{\Omega_\mathrm{B}(a-r)^2},
\end{equation}
where $v_\mathrm{B}$, $\Omega_\mathrm{B}$, and $\dot{M}_\mathrm{B}$
are respectively the velocity, solid angle and rate of
the mass outflow from the Be star and $a$ is the orbital separation
of the binary system (see \cite{melatos1995}
for the more actual shape of the shock).
Then, the shock distance $r_\mathrm{s}$ is
\begin{equation}
	r_\mathrm{s} = \xi \cdot a
\label{eq:shock_distance}
\end{equation}
where $\xi$ is given by
\begin{equation}
	\xi^{-1} = \sqrt{\frac{\Omega_\mathrm{w}}{\Omega_\mathrm{B}} \cdot
	\frac{\dot{M}_\mathrm{B}v_\mathrm{B}c}{\dot{E}_\mathrm{rot}}}+1.
\end{equation}
If the pulsar and stellar winds are spherically symmetric,
$\xi$ remains constant throughout the orbit,
no matter how eccentric it is.
In the PSR B1259-63/SS 2883 system the orbital separation
and hence the shock distance
may vary by an order of magnitude between periastron and apastron
because of the deformed binary orbit with eccentricity of $e=0.86$.
Note that the shock distance depends also on the physical parameters
and geometry of pulsar and Be star winds, which are not so certain.
Hence, in the following we treat $\xi$ as a free parameter
which can be determined from the comparison with observations.

The shock in the pulsar wind compresses the magnetic field as well as
the wind matter.
The field strength $B_2$ at the downstream side of the shock is
calculated from equation (\ref{eq:mag1}) together with
the relativistic MHD shock conditions \citep{ken84},
\begin{equation}
	B_2(r_\mathrm{s}) = 3B_1(r_\mathrm{s}).
\label{eq:mag2}
\end{equation}

The shock accelerates the pulsar wind particles,
electrons and positrons, to the relativistic high energies
\citep{hoshi92}.
We assume that the acceleration time is short compared to
the radiative cooling times and the flow time
of the shock downstream region.
We adopt the power law distribution for the accelerated
electrons and positrons in the range of $\gamma_1 < \gamma < \gamma_2$
\begin{equation}
	N_0(\gamma) = K \gamma^{-p},
\label{eq:nonthermal_density}
\end{equation}
where $N_0(\gamma)$ is the density
of relativistic electrons and positrons injected at the shock,
$\gamma$ the Lorentz factor, $p$ the power law index
and $K$ the constant.
We assume that the distributions of accelerated particles at
the shock are the same and given by equation
(\ref{eq:nonthermal_density}) irrespective of the binary phase.
A fraction $\varepsilon_\mathrm{a}$ of the pulsar spin-down luminosity
is transformed into the relativistic electrons and positrons.
The constant $K$ in equation (\ref{eq:nonthermal_density}) is
then given by
\begin{equation}
	K = (p-2){\gamma_1}^{p-1}
		\frac{\varepsilon_\mathrm{a}\dot{E}_\mathrm{rot}}
			{\Omega_\mathrm{w}{r_\mathrm{s}}^2\gamma_1 mc^2(c/3)}
\label{eq:K_const}
\end{equation}
for $p \neq 2$.
In equation (\ref{eq:K_const}) $m$ is the electron mass.
The accelerated relativistic particles are injected into
the downstream region at the shock and flow
in the downstream region with a speed of $c/3$.

\section{Cooling Times}
\label{sec:cooling_times}
Relativistic electrons and positrons lose energy
through the synchrotron and inverse Compton processes,
emitting X-rays and $\gamma$-rays,
as they flow along the stream line in the postshock region.
Relativistic electrons and positrons may also suffer from the
adiabatic energy loss.
Here we estimate the cooling times of the relativistic electrons and positrons
using the physical parameters at the shock.

The flow time of the postshock stream is given approximately by
\begin{equation}
	\tau_\mathrm{f} = \frac{r_\mathrm{s}}{c/3}.
\label{eq:flow_time}
\end{equation}
The flow time $\tau_\mathrm{f}$ denotes approximately the residence
time of relativistic electrons and positrons in the emission nebula.
The flow time $\tau_\mathrm{f}$ also represents the cooling time
of the relativistic particles due to the adiabatic energy loss.
Note that the power law index of the energy distribution
of relativistic electrons and positrons remains
unchanged throughout adiabatic cooling.

The typical energy of synchrotron photons is
\begin{equation}
	\varepsilon \sim 0.23\hbar
		\left(\frac{3\gamma^2 e B_2}{2mc}\right),
\end{equation}
where $\varepsilon$ is the photon energy,
$e$ the electron charge and $\hbar$ the Planck constant.
The synchrotron cooling time is calculated by
\begin{equation}
	\tau_\mathrm{s} =
		\frac{3mc}{4\sigma_\mathrm{T}U_\mathrm{B_2}\gamma},
\end{equation}
where $\sigma_\mathrm{T}$ is the Thomson cross section and
$U_\mathrm{B_2}$ is the magnetic energy density,
\begin{equation}
	U_\mathrm{B_2}(r_\mathrm{s}) = \frac{{B_2}^2}{8\pi}.
\end{equation}
The relativistic electrons and positrons lose energy by scattering off
the photons emitted from the Be star.
The typical energy of the scattered photons is
\begin{equation}
	\varepsilon \sim
	\left\{
		\begin{array}{rl}
		\gamma^2 \varepsilon_0 & (\gamma\varepsilon_0 \ll mc^2) \\
		\gamma mc^2 & (\gamma\varepsilon_0 \gg mc^2)
		\end{array}
	\right. ,
\label{eq:scattering_energy}
\end{equation}
where $\varepsilon_0$ is the typical energy of photons
from the Be star.
In the Thomson limit, where $\gamma\varepsilon_0 \ll mc^2$,
the inverse Compton cooling time is calculated by
\begin{equation}
	\tau_\mathrm{ic} =
		\frac{3mc}{4\sigma_\mathrm{T}U_\mathrm{p}\gamma},
\label{eq:cooling_time_ic_Thomson}
\end{equation}
where $U_\mathrm{p}$ is the energy density of photons from the Be star.
The photon energy density $U_\mathrm{p}$ at the shock is estimated by
\begin{equation}
	U_\mathrm{p}(r_\mathrm{s}) =
		\frac{{R_\mathrm{B}}^2\sigma_\mathrm{B}T^4}
		{(a-r_\mathrm{s})^2 c},
\end{equation}
where $R_\mathrm{B}$ is the radius of the Be star,
$T$ the effective surface temperature and $\sigma_\mathrm{B}$
the Stefan-Boltzman constant.
In the Klein-Nishina limit,
where $\gamma\varepsilon_0 \gg mc^2$,
the inverse Compton scattering becomes less efficient and furthermore,
a large fraction of the electron (positron)
energy can be lost in one scattering.
Hence, in general,
we need to compute the average energy loss rate
$\langle -\dot{\gamma}_\mathrm{ic}\rangle$
using the Klein-Nishina formula, in which case
\begin{equation}
	\langle -\dot{\gamma}_\mathrm{ic}\rangle =
		\int^\gamma_1
			P(\gamma,\gamma')(\gamma-\gamma')\mathrm{d}\gamma',
\label{eq:energy_loss_KN}
\end{equation}
where $P(\gamma,\gamma')\mathrm{d}\gamma'\mathrm{d}t$
is the probability that an electron
with energy $\gamma$ will undergo a collision causing it to fall down
to the lower energy state between $\gamma'$
and $\gamma'+\mathrm{d}\gamma'$ in time $\mathrm{d}t$.
Here we calculate the probability $P(\gamma,\gamma')$
for a blackbody photon field.
For details of $P(\gamma,\gamma')$ we refer the readers
to \citet{jon68} and \citet{blu70}.
Then, the inverse Compton cooling time is redefined by
\begin{equation}
	\tau_\mathrm{ic} =
		\frac{\gamma}{\langle -\dot{\gamma}_\mathrm{ic}\rangle}.
\label{eq:cooling_time_ic_general}
\end{equation}
Note that equation (\ref{eq:cooling_time_ic_general})
reduces to equation ({\ref{eq:cooling_time_ic_Thomson})
in the Thomson limit.

\begin{figure}
\begin{center}
	\FigureFile(83mm,58.6mm){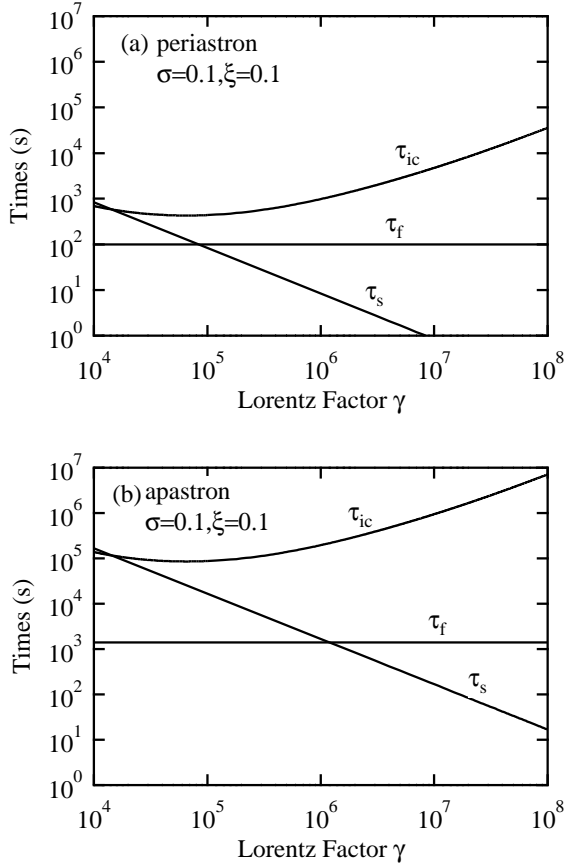}
	\caption{Cooling and flow times of relativistic electrons
		and positrons at periastron (a) and apastron (b) plotted
		against the Lorentz factor.
		$\tau_\mathrm{s}$, $\tau_\mathrm{ic}$ and $\tau_\mathrm{f}$
		express the synchrotron cooling time, the inverse Compton
		cooling time and the flow time, respectively.
		The model parameters used are $\sigma=0.1$ and $\xi=0.1$.
		The synchrotron cooling dominates over the inverse Compton
		cooling. The increase of $\tau_\mathrm{ic}$ for larger
		$\gamma$ reflects the Klein-Nishina effect.}
	\label{fig:1}
\end{center}
\end{figure}
We illustrate the cooling times at periastron and apastron
as a function of the Lorentz factor
of relativistic electrons and positrons in figure \ref{fig:1}.
The binary separations at periastron and apastron
are $a=9.9 \times 10^{12}\ \mathrm{cm}$
and $a=1.4 \times 10^{14}\ \mathrm{cm}$,
respectively \citep{manche95}.
The spin down luminosity of the pulsar is
$\dot{E}_\mathrm{rot} \sim 8.28 \times 10^{35}\ \mathrm{erg/s}$.
We take the solid angle of the pulsar wind
to be $\Omega_\mathrm{w} \sim \pi$.
We adopt $R_\mathrm{B} \sim 11\RO$ and
$T \sim 2.7 \times 10^4\ \mathrm{K}$
for the radius and effective surface
temperature of the Be star \citep{johns92b}.
The physical parameters $\sigma$ and $\xi$ are chosen as
$\sigma=0.1$ and $\xi=0.1$.
The field strengths at the shock are $B_2 \sim 9.6\ \mathrm{G}$
and $B_2 \sim 0.67\ \mathrm{G}$ for periastron
(figure \ref{fig:1}a) and apastron (figure \ref{fig:1}b), respectively.
The synchrotron cooling time decreases simply with
increasing Lorentz factor.
On the other hand, the inverse Compton cooling time first
decreases and then increases with increasing Lorentz factor.
This behavior is a consequence of the Klein-Nishina effect
that the inverse Compton scattering becomes less
efficient at higher energies.
The synchrotron emission dominates over the inverse
Compton scattering at both periastron and apastron
for $\gamma \gtrsim 10^5$, which approximately corresponds
to synchrotron emission in the X-ray band.
Both the synchrotron and inverse Compton cooling times
are longer at apastron than at periastron since the field
strength and photon energy density are lower at apastron.
Note that the synchrotron cooling time becomes shorter than
the flow time at $\gamma > \gamma_\mathrm{b} \sim 10^5$
and at $\gamma > \gamma_\mathrm{b} \sim 10^6$
in figure \ref{fig:1}a and figure \ref{fig:1}b, respectively.
As shown in section \ref{sec:nebula_emission},
there appears a break-off in the slope of the particle
distribution and emission spectrum at the energy
corresponding to $\gamma_\mathrm{b}$
since higher energy particles are lost rapidly.

\begin{figure}
\begin{center}
	\FigureFile(83mm,59.4mm){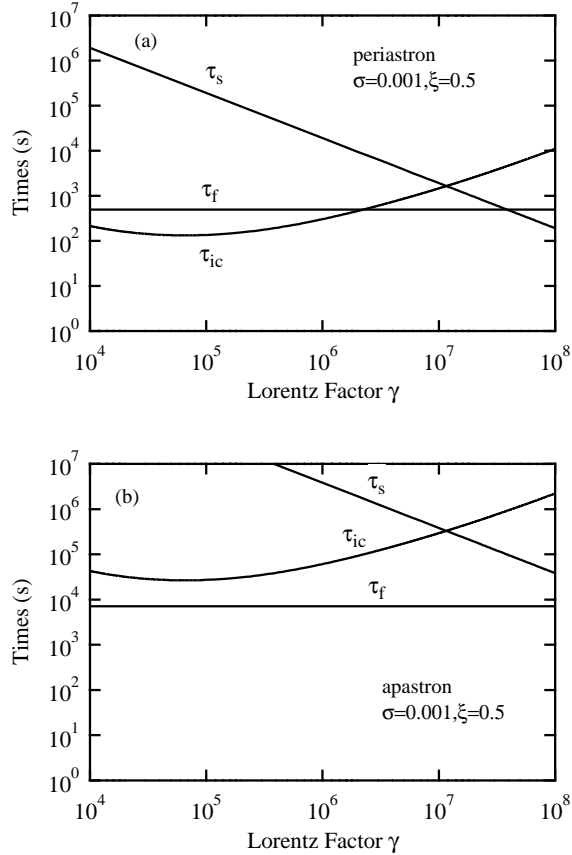}
	\caption{Cooling and flow times of relativistic electrons
		and positrons at periastron (a) and apastron (b) for
		the case with $\sigma=0.001$ and $\xi=0.5$.
		The inverse Compton cooling dominates over
		the synchrotron cooling.}
	\label{fig:2}
\end{center}
\end{figure}
We depict the cooling times also for the parameters
$\sigma=0.001$ and $\xi=0.5$ in figure \ref{fig:2}.
The field strengths at the shock are $B_2 \sim 0.2\ \mathrm{G}$
and $B_2 \sim 0.01\ \mathrm{G}$ for periastron
(figure \ref{fig:2}a) and apastron (figure \ref{fig:2}b), respectively.
The magnetic fields adopted are
considerably weaker than those in figure \ref{fig:1}.
The energy densities of photons are higher than
those in figure \ref{fig:1} since the shock is located
closer to the Be star.
The inverse Compton scattering dominates over the synchrotron
emission and hence determines the energy distribution
of relativistic electrons and positrons in the emission nebula.

\section{Nebula Emission}
\label{sec:nebula_emission}

\begin{figure}
\begin{center}
	\FigureFile(83mm,59.4mm){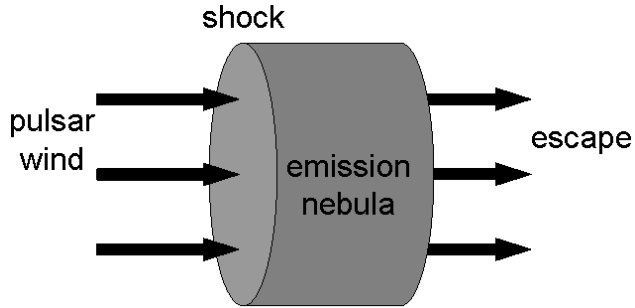}
	\caption{Schematic diagram of an emission nebula.}
	\label{fig:3}
\end{center}
\end{figure}

An emission nebula, which radiates X-rays and $\gamma$-rays,
is formed in the shock downstream region.
The geometry of the emission nebula depends on the properties
of the pulsar and Be star winds.
Here, for simplicity, we approximate the nebula geometry
by a cylinder that has the cross section of
$\Omega_\mathrm{w}{r_\mathrm{s}}^2$
and the length of $L \sim r_\mathrm{s}$ (figure \ref{fig:3}).
The accelerated electrons and positrons are injected
at the termination shock (one end of the cylinder)
and flow in the direction of the central axis
of a cylinder together with the shocked pulsar wind,
keeping the density constant.
Relativistic electrons and positrons emit
X-rays and $\gamma$-rays through the synchrotron
and inverse Compton processes.
At the other end of the cylinder they escape
out of the emission nebula.

\subsection{Distribution of
	Relativistic Electrons and Positrons}
The energy distribution of relativistic electrons and
positrons varies as they flow away from the shock,
following a streamline, because of the radiative energy loss.
The evolution of the particle distribution $N(\gamma,t)$
can be described by the integro-differential equation,
\begin{eqnarray}
	\lefteqn{\frac{\partial N(\gamma,t)}{\partial t}
	+ \frac{\partial}{\partial\gamma}
		\left[\dot{\gamma}N(\gamma,t)\right]}\hspace{.4cm} \nonumber \\
	&+& N(\gamma,t)\int^\gamma_1 \mathrm{d}\gamma'P(\gamma,\gamma')
	- \int^\infty_\gamma \mathrm{d}\gamma'
		N(\gamma',t)P(\gamma',\gamma) \nonumber \\
	& & \qquad\qquad\qquad
		 = N(\gamma,0)\delta(t)-\frac{N(\gamma,t)}{\tau_\mathrm{f}},
\label{eq:particle_distribution}
\end{eqnarray}
where $t$ is the time that elapsed after the injection
of relativistic particles at the shock and is related to the
distance $x$ from the shock by $x=(c/3)t$.
$\dot{\gamma}$ in equation (\ref{eq:particle_distribution}) 
represents the energy loss due to the synchrotron process,
while the third and fourth terms in the left-hand side represent
the energy loss due to the inverse Compton process.
The first term in the right-hand side of
equation~(\ref{eq:particle_distribution}) represents the source
of the relativistic electrons and positrons
(equation~(\ref{eq:nonthermal_density})),
supplied at $t=0$ (or at the shock), while the second term
represents their escape from the emission nebula.
The adiabatic energy loss is ignored.

\begin{figure}
\begin{center}
	\FigureFile(83mm,59.3mm){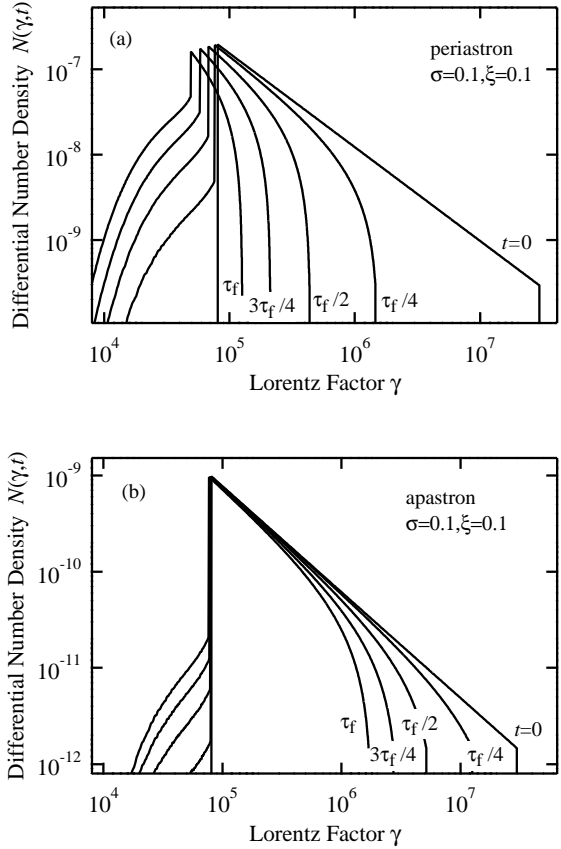}
	\caption{Variation of the energy distribution of relativistic
		electrons and positrons with time (or with distance from
		the shock) at periastron (a) and apastron (b).
		The parameters for the magnetization and shock position
		are taken as $\sigma=0.1$ and $\xi=0.1$.
		The power law function, characterized by the index
		of $p=2.1$, the minimum Lorentz factor of
		$\gamma_1 = 8 \times 10^4$, and the maximum Lorentz factor
		of $\gamma_2 = 3 \times 10^7$, is used as the
		input particle distribution. The particle distributions
		at $t=0$, $\tau_\mathrm{f}/4$, $\tau_\mathrm{f}/2$,
		$3\tau_\mathrm{f}/4$ and $\tau_\mathrm{f}$ are illustrated.
		The particles are lost from higher energies with time
		due to the radiative loss, especially synchrotron loss,
		and accumulate at lower energies.
		The variation of the particle distribution at periastron
		is much larger than that at apastron because of
		the efficient synchrotron cooling.}
	\label{fig:4}
\end{center}
\end{figure}
We show in figure \ref{fig:4} how the energy distribution
of relativistic electrons and positrons changes
with time or with distance
from the shock for the periastron and apastron cases.
Here the parameters for the magnetization and shock position are
taken as $\sigma=0.1$ and $\xi=0.1$,
while the parameters for the input particle
distribution are taken as
$p=2.1$, $\gamma_1=8 \times 10^4 $, and $\gamma_2=3 \times 10^7 $.
The radiative loss is mainly determined by the synchrotron process.
The loss of relativistic electrons and positrons
occurs faster at higher Lorentz factors
and extends to the lower Lorentz factors as with time.
The particles accumulate at $\gamma<\gamma_1$ with time
since the particles fall from higher
to lower Lorentz factors losing energy.
The particle distribution undergoes a large variation with
time due to the efficient synchrotron cooling in the periatron case
(figure \ref{fig:4}a), whereas in the apastron case
(figure \ref{fig:4}b) the variation is less significant.

\begin{figure}
\begin{center}
	\FigureFile(83mm,107.5mm){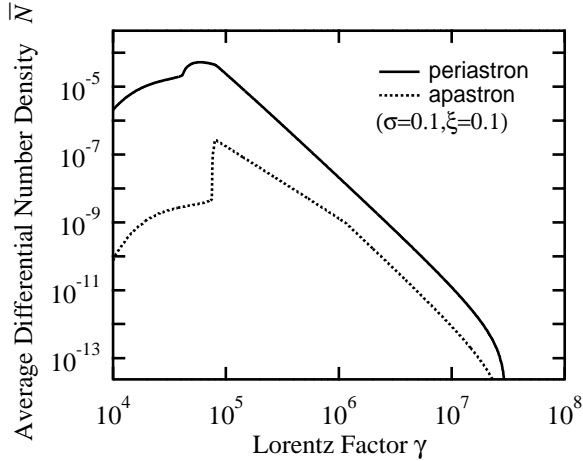}
		\caption{Average distributions of relativistic electrons and
		positrons in the nebula at periastron (solid line) and
		at apastron (dotted line). The model parameters used are
		the same as those in figure \ref{fig:4}.
		The synchrotron cooling is dominant over the inverse Compton
		cooling. The slope of the particle distribution steepens
		by one, compared to the input particle distribution,
		above $\gamma_\mathrm{b} \sim 10^5$ and
		$\gamma_\mathrm{b} \sim 10^6$, where the synchrotron cooling
		time becomes equal to the flow time (figure \ref{fig:1}),
		respectively for the periastron and apastron cases.}
	\label{fig:5}
\end{center}
\end{figure}
We integrate the relativistic electrons and positrons,
shown in figure \ref{fig:4},
over an entire emission nebula and then divide the integrated
one by the nebula volume.
We obtain the average energy distribution
for the relativistic electrons and positrons in the nebula,
which we show in figure \ref{fig:5}.
As expected from figure~\ref{fig:1},
we observe that the slope of the particle distribution steepens
by one, compared to the input particle distribution,
above $\gamma_\mathrm{b} \sim 10^5$ and $\gamma_\mathrm{b} \sim 10^6$,
where the synchrotron cooling time
becomes shorter than the flow time (figure \ref{fig:1}),
in the periastron and apastron cases, respectively.
Note that the total number of relativistic electrons and
positrons in the emission
nebula is larger at apastron than at periastron,
contrary to the particle density, since the volume of
the emission nebula is assumed to be in proportion to
${r_\mathrm{s}}^3$.

\begin{figure}
\begin{center}
	\FigureFile(83mm,117.35mm){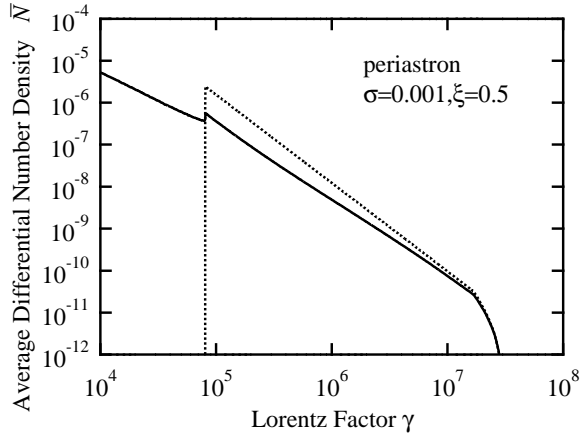}
	\caption{Average distribution of relativistic electrons and positrons
		at periastron for the case with $\sigma=0.001$ and $\xi=0.5$.
		The inverse Compton cooling is dominant over the synchrotron
		cooling. In the solid line both the inverse Compton and
		synchrotron cooling are included, while in the dotted line
		the inverse Compton cooling is suppressed completely.
		The inverse Compton cooling causes flattening in
		the slope of the particle distribution.}
	\label{fig:6}
\end{center}
\end{figure}
Next, let us consider the average energy distribution of
relativistic electrons and positrons when the energy loss of particles
is determined mainly by the inverse Compton process.
Adopting the same values for parameters $\sigma$ and $\xi$
as those in figure \ref{fig:2}, we plot the average
distribution at periastron in figure \ref{fig:6}.
In the solid line we include both the inverse Compton and synchrotron
cooling although the inverse Compton cooling is far dominant over
the synchrotron cooling.
In the dotted line we retain only the synchrotron cooling,
suppressing the inverse Compton cooling completely,
in order to understand the effect of the inverse Compton process
on the particle distribution.
We observe in the solid line the flattening rather than
steepening for the slope of the particle distribution at
$\gamma > 10^5$, contrary to the synchrotron dominant case
in figure \ref{fig:5}.
This flattening is caused by the Klein-Nishina effect.
As seen from figure \ref{fig:2},
the inverse Compton cooling is most efficient at $\gamma \sim 10^5$
and then becomes less efficient at $\gamma > 10^5$
because of the Klein-Nishina effect.
The increase of the cooling time with increasing $\gamma$
yields the flattening in the particle distribution.

\subsection{Radiation Spectrum}
The relativistic electrons and positrons in the nebula
radiate $\mathrm{keV\!-\!MeV}$ photons
through the synchrotron process and
$\mathrm{GeV\!-\!TeV}$ photons
through the inverse Compton process.
We calculate the spectrum of X-rays and $\gamma$-rays
radiated from the whole nebula,
taking fully into account the particle distribution,
the Klein-Nishina effect, the Planck distribution for
the target photons of the inverse Compton scattering
and the spectral function for synchrotron emission.
We illustrate the X-ray and $\gamma$-ray spectra expected
at periastron and apastron in figure \ref{fig:7},
where we adopt the same particle distribution and physical parameters
as those used in figure \ref{fig:5}.
\begin{figure}
\begin{center}
	\FigureFile(83mm,110.7mm){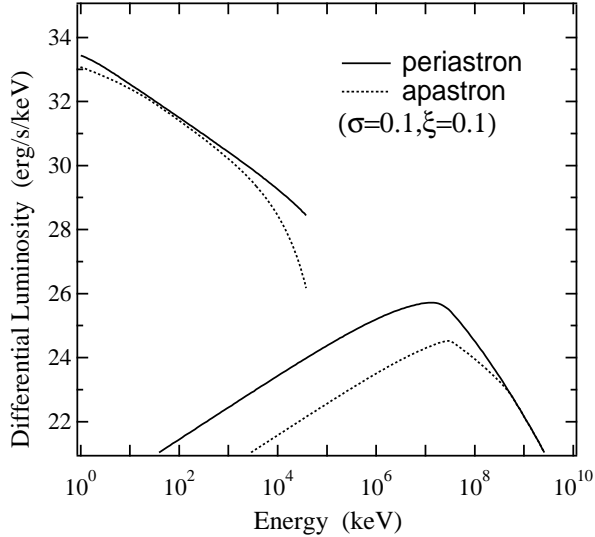}
	\caption{Spectra of nebula emission at periastron (solid line)
		and apastron (dotted line). $\mathrm{KeV\!-\!MeV}$ photons
		are emitted by the synchrotron process, while
		$\mathrm{GeV\!-\!TeV}$ photons by the inverse Compton process.
		The model parameters used are the same as
		those in figure \ref{fig:5}.}
	\label{fig:7}
\end{center}
\end{figure}

The spectral shape in the X-ray band is well represented by
a power law function with exponents of $\sim 1$ and $\sim 0.7$
at periastron and apastron, respectively.
The steeper gradient at periastron is a direct consequence of
a steeper particle distribution at periastron,
caused by the efficient synchrotron cooling.
The X-ray luminosity decreases slightly with the binary separation.
This weak dependence on the binary separation of the X-ray luminosity
can easily be understood by using equations
(\ref{eq:mag1}),
(\ref{eq:shock_distance}),
(\ref{eq:mag2}) and
(\ref{eq:nonthermal_density}).
Physically, the insensitivity of the X-ray luminosity arises
because the injected energy remains constant over the orbit
and this is converted almost completely into X-rays.
If we adopt the larger values for the maximum Lorentz factor
$\gamma_2$, the spectrum of the synchrotron component
extends to higher energies.
Hence, if the sensitivity of observations in the
$10\ \mathrm{MeV}\!-\! 1\ \mathrm{GeV}$ band is improved in future,
the maximum Lorentz factor $\gamma_2$ of the particle
acceleration would be determined.

The inverse Compton component peaks at a $\gamma$-ray energy of
$\sim 10^{10}\ \mathrm{eV}$,
as expected from equation (\ref{eq:scattering_energy})
together with the
Lorentz factor at the peak of the particle distribution.
Hence, the minimum of the input Lorentz factor, which may lie near
the peak of the particle distribution, can be estimated
from the observed $\gamma$-ray energy at the peak.
The peak height of the inverse Compton component decreases
significantly with binary separation.
We note that the density of target photons decreases in
inverse proportion to the square of binary separation,
while the number of relativistic electrons and positrons responsible
for the peak in the whole nebula is determined by the flow time
(or the escape time) and hence increases in proportion
to the binary separation.
In consequence, the inverse Compton scattering over the whole nebula
becomes less frequent as the binary separation enlarges.
The spectral curves, as seen from the periastron and apastron cases,
converge and become independent of the binary
separation near the highest end of $\gamma$-ray energies.
We should note that the number of
ultrarelativistic electrons and positrons
responsible for these high energy $\gamma$-rays is
determined by the synchrotron process and hence is inversely
proportional to the energy density of the magnetic field
whereas the inverse Compton scattering rate is proportional
to the energy density of target photons.
For constant $\xi$ the ratio of magnetic energy density to
photon energy density at the shock stays constant
throughout the orbit, which results in the independence of
spectral curves on the binary separation.

If the magnetic field at the shock is weak and the inverse
Compton cooling dominates the synchrotron cooling,
a larger number of relativistic electrons and positrons is required to
produce such X-ray luminosity as shown in figure \ref{fig:7}.
Then, the $\mathrm{GeV\!-\!TeV}$ $\gamma$-ray luminosity is enhanced
considerably compared to that in figure \ref{fig:7} \citep{kirk1999a}.

The adiabatic loss, even if included in calculations,
yields only the slight change on the spectral curve in
fiugre \ref{fig:7} and may not alter the essence of our results.

The target photons at the nebula are
assumed here to be isotropic
in computing the inverse Compton emission.
In the case of photons from the Be star of a binary companion,
the distribution of target photons may be closer to
unidirectional rather than isotropic.
Adopting the unidirectional distribution, \citet{kirk1999a}
calculated the inverse Compton emission and show that
the inverse Compton emission depends not only on
the binary separation, but also on the binary orientation.
We estimate from comparison of our computed spectra with
their results that the effects of target photon distribution
may produce a change of less than a factor of a few in
the high energy $\gamma$-ray spectrum of inverse Compton emission.
Hence, the $\gamma$-ray spectra of inverse Compton component
shown in fiugre \ref{fig:7} is approximate.
The effects of binary orientation should be included in
the model especially when we conduct the closer comparison
with the high energy $\gamma$-ray observations of
PSR 1259-63/Be star system.

The compression ratio, used in equations (\ref{eq:mag2}),
(\ref{eq:K_const}) and (\ref{eq:flow_time}),
is fixed to 3 throughout the paper although it depends on
the Lorentz factor and magnetization parameter of
the upstream flow \citep{ken84,kirk1999b}.
The value of 3 is appropriate when the upstream flow is
highly relativistic and the magnetization parameter is very small.
The compression ratio is $\sim 2.6$ for
$\gamma_\mathrm{w} \sim 10^5$ and $\sigma \sim 0.1$
used in figure \ref{fig:7}.
This change in compression ratio almost does not
alter the synchrotron component in figure \ref{fig:7}
while it slightly modifies the inverse Compton component.

\section{Comparison with Observations}
\label{sec:comparison}
The soft and hard X-rays observed from the PSR B1259-63/SS 2883 system
have the following characteristics \citep{hira99}: (1) the X-ray
spectra are represented by a power law function that extends
from 1 to 200 keV; (2) the spectral index varies with orbital phase,
from $\sim 2$ at periastron to $\sim 1.6$ at apastron; (3) the X-ray
luminosity in the $1\!-\!10\ \mathrm{keV}$
band varies with orbital phase
by about an order of magnitude, from $\sim 10^{34}\ \mathrm{ergs/s}$
near periastron to $\sim 10^{33}\ \mathrm{ergs/s}$ at apastron,
while just at periastron the light curve displays a drop by a factor
of two compared to that slightly before or after the periastron;
(4) the pulsation is absent in the X-ray time series.

\begin{figure}
\begin{center}
	\FigureFile(175mm,254.8mm){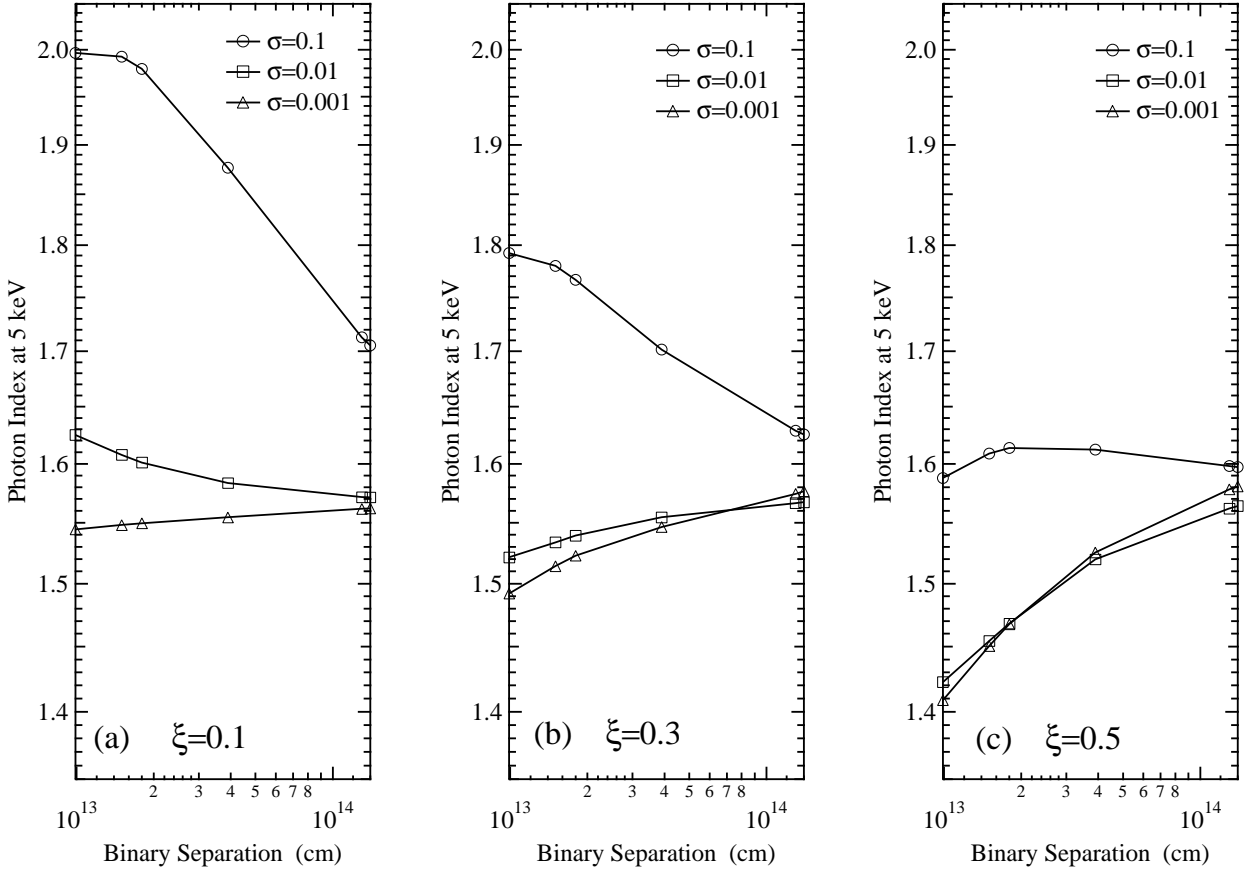}
	\caption{Photon spectral indices calculated at 5 keV for
		various model parameters as a function of the binary
		separation. For larger $\sigma$ and smaller $\xi$ the
		synchrotron emission dominates over the inverse Compton
		scattering, leading to the spectral steepening.
		For smaller $\sigma$ and larger $\xi$, on the other hand,
		the inverse Compton scattering dominates over the
		synchrotron emission, leading to the spectral flattening.}
	\label{fig:8}
\end{center}
\end{figure}
We plot the photon spectral gradient (index) calculated at
$5\ \mathrm{keV}$ for various model parameters as a function of
the binary separation in figure \ref{fig:8}.
The shock positions in figures \ref{fig:8}b and \ref{fig:8}c
are taken to be 3 and 5 times farther away from the pulsar,
respectively,  compared to that in figure \ref{fig:8}a.
The magnetic field strengths at the shock are larger for
larger $\sigma$ and smaller $a$ and $\xi$,
while the target photon densities are larger
for larger $\xi$ and smaller $a$.
We assume that the distributions of
relativistic electrons and positrons injected
at the shock are the same irrespective of the binary phase.
We fix the minimum and maximum Lorentz factors for the input
particle distribution to $\gamma_1 \sim 8 \times 10^4$
and $\gamma_2 \sim 3 \times 10^7$, respectively.
The change of the particle distribution in the nebula due to
the radiative cooling is not significant at apastron.
Hence, the photon spectral index at apastron reflects the original
slope of the accelerated particle distribution.
We adopt $p \sim 2.1$ determined from the photon index
of $\sim 1.6$ observed at apastron.
The observed photon index variation with orbital phase
is approximately reproduced by the case with $\sigma \sim 0.1$
and $\xi \sim 0.1$ in figure \ref{fig:8}a.
Contrary to the results in figure 11 of \citet{tav97},
we find that the efficient synchrotron cooling is responsible
for the spectral steepening in the X-ray band around periastron.
As clearly seen at periastron for $\sigma \sim 0.001$
in figures \ref{fig:8}b and \ref{fig:8}c,
if the inverse Compton cooling dominates,
the spectral slope rather flattens because of the Klein-Nishina effect.
If the Klein-Nishina effect is fully taken into account,
some results of \citet{tav97} may be altered.

\begin{figure}
\begin{center}
	\FigureFile(175mm,254.8mm){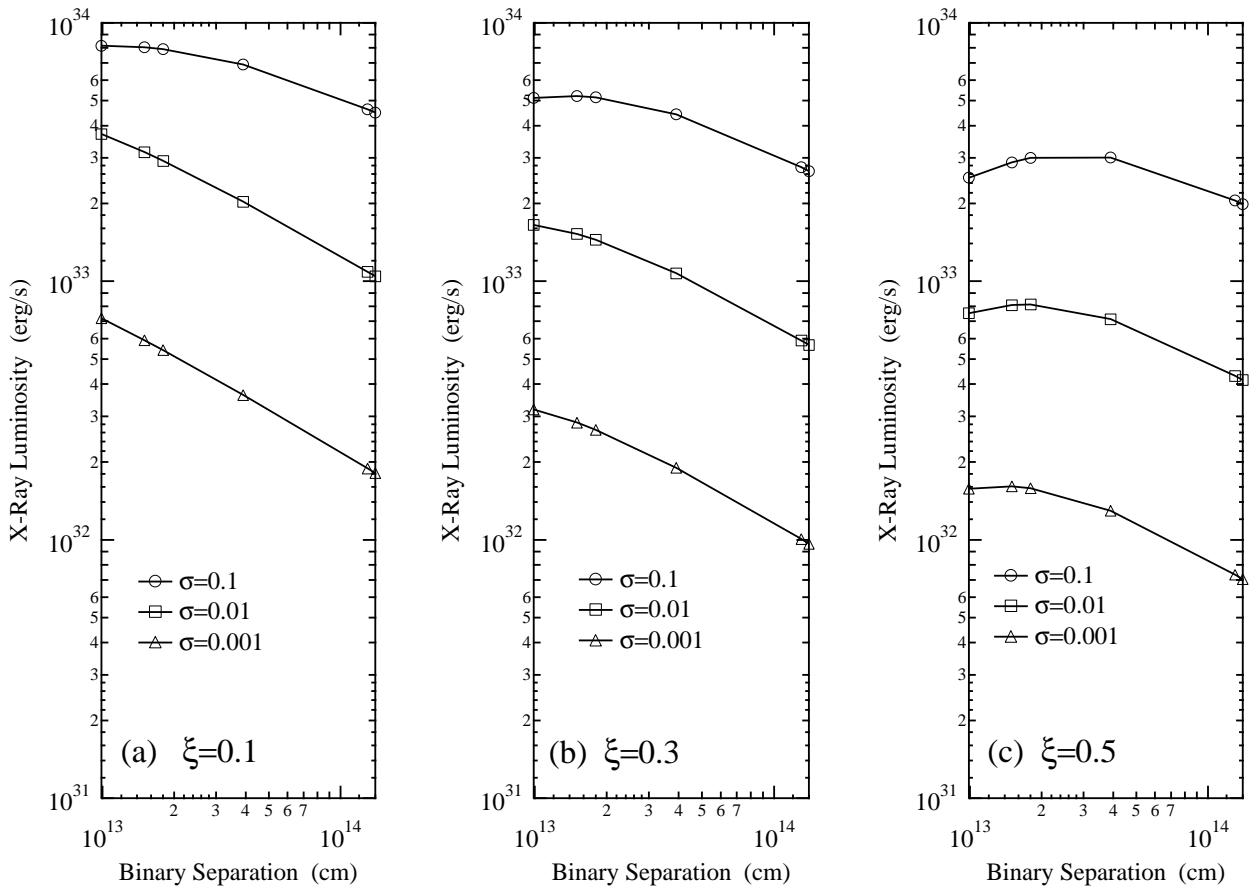}
	\caption{Luminosity of X-rays in the $1\!-\!10\mathrm{keV}$
		band calculated for various model parameters as
		a function of the orbital separation.
		X-ray photons are emitted by the synchrotron process.
		The magnetic field strength in the nebula and hence the
		synchrotron emission rate decrease with decreasing $\sigma$.}
	\label{fig:9}
\end{center}
\end{figure}
If we choose $\xi \sim 0.1$ and $\sigma \sim 0.1$,
the energy conversion parameter $\varepsilon_\mathrm{a}$ is determined
to be $\sim 0.1$ from reproducing the relatively large X-ray
luminosity observed at periastron.
We depict the X-ray luminosity in the $1\!-\!10\ \mathrm{keV}$
band calculated for $\varepsilon_\mathrm{a} \sim 0.1$ and various
other model parameters as a function of
the binary separation in figure \ref{fig:9}.
Note that, if the model parameters $\xi$ and $\sigma$
are taken to be constant throughout the binary orbit,
the luminosity variation with orbital
separation predicted by the model is too small to explain
the observed variation amounting to an order of magnitude.
We point out that this luminosity problem may be resolved
if the magnetization parameter $\sigma$ decreases with
orbital separation and the shock position parameter
$\xi$ increases with orbital separation.

The magnetization parameter is likely to decrease with distance
from the neutron star in the pulsar wind.
Note that the small value of $\sim 0.003$ is reported for
the Crab pulsar wind from the nebula analysis \citep{ken84}.
In the Crab pulsar the distance of the shock,
where the magnetization parameter is diagnosed,
from the neutron star is $\sim 0.1\ \mathrm{pc}$,
while in the PSR B1259-63 $\sim 10^{12}\!-\!10^{13}\ \mathrm{cm}$.
When combined with our result of the PSR B1259-63,
the result of the Crab pulsar strengthens the conclusion that
the magnetization parameter may decrease with distance
from the neutron star.
The magnetization parameter and shock distance presented
here can be used to study the property of a pulsar wind.

Our results shown in figure \ref{fig:7} are consistent with
the COMPTEL and EGRET upper limits
in the $1\!-\!1000\ \mathrm{MeV}$ range.
Recently, the CANGAROO collaboration reports the new upper limit
on the emission in the TeV range \citep{kawa02},
which is well above our prediction and consistent with our results.
Note, however, that the reproduction of a luminosity drop in X-rays,
observed just at periastron, is beyond the scope of our simple model.
We need to construct the detailed model that
includes the binary orientation,
the flow patterns of the Be star and pulsar winds,
the misalignment of the pulsar orbital plane with the Be star
outflow disk, the shielding of the emission region from
the target photons for the inverse Compton scattering and
the adiabatic loss in addition to the variation of model
parameters $\sigma$ and $\xi$ with orbital separation.

\section{Concluding Remarks}
\label{sec:conclusion}
The PSR 1259-63/Be star system can provide important information
on the pulsar wind.
We can diagnose the property of the pulsar wind at various
distances from the pulsar by applying the shock powered
emission model to the X-ray fluxes and spectra observed
at different orbital phases because
the binary orbit is highly eccentric.
We find that the magnetization parameter of the pulsar wind,
the ratio of the Poynting flux to the kinetic energy flux,
is $\sim 0.1$ at the distance of $\sim 10^{12}\ \mathrm{cm}$
and may decrease with distance from the pulsar.
Note that the magnetization parameter of the Crab pulsar wind
is $\sim 0.003$ at the distance of $\sim 0.1\ \mathrm{pc}$.
These two results, when combined, strengthen our conclusion
that the magnetization parameter may decrease
with distance from the neutron star.
We need more X-ray data from various pulsar nebulae
in order to determine the pulsar wind property as a function
of distance from the pulsar and to understand the energy conversion
from the Poynting flux to the kinetic energy flux in the pulsar wind.

The relativistic electrons and
positrons of pulsar winds undergo
inverse Compton scattering also upstream of the termination shock
\citep{chernyakova1999,bogovalov2000,ball2000}.
\citet{ball2000} show that $\gamma$-rays resulting from
the inverse Compton scattering in the wind of
PSR 1259-63 may be detectable by atmospheric
Cerenkov detectors or by the programs
INTEGRAL and GLAST if the size of the pulsar wind nebula
is comparable to the binary separation.
Then, we can diagnose directly the unshocked regions
of pulsar winds previously thought to be invisible.

The PSR 1259-63/Be star system can also provide valuable
information on the shock acceleration.
We find that a fraction of $\sim 0.1$ of the pulsar energy
is transformed into the relativistic electrons and positrons.
We obtain the power law index of $\sim 2$ for the distribution
of relativistic electrons and positrons injected at the shock.
We should note that the maximum energy of
accelerated electrons and positrons
can be estimated from the break-off energy in the photon spectrum.
We expect this spectral break in the $\mathrm{MeV\!-\!GeV}$ range,
which may be detectable by the future program GLAST.
We may derive the minimum energy of
relativistic electrons and positrons
from the high energy $\gamma$-rays due to the inverse Compton emission.
The observations of $\mathrm{GeV\!-\!TeV}$
$\gamma$-rays are also encouraged.

\bigskip
We thank M.~Hirayama, F.~Takahara and M.~Hoshino for
useful discussions and comments.
We also thank the anonymous referee for valuable comments.
This work was partly supported by a
grant-in-aid of Ministry of Education, Culture,
Sports, Science and Technology (12640302).

\end{document}